\documentclass[a4paper,11pt,dvips,colordvi]{article}
\usepackage{amssymb}
\usepackage{amsmath}
\usepackage{color}
\usepackage{multirow}

\usepackage{colordvi}

\newcommand{\beq}{\begin{equation}}
\newcommand{\eeq}{\end{equation}}
\newcommand{\beqn}{\begin{eqnarray}}
\newcommand{\eeqn}{\end{eqnarray}}

\def\av#1{\langle #1 \rangle}

\newcommand{\om}{\omega}

\newcommand{\si}{\sigma}

\newcommand{\te}{\theta}

\newcommand{\pa}{\partial}
\newcommand{\al}{\alpha}

\newcommand{\lga}{\longrightarrow}

\newcommand{\ka}{\kappa}

\newcommand{\da}{\dagger}

\newcommand{\NP}[1]{ {\it Nucl.~Phys.} {\bf #1}}

\newcommand{\PR}[1]{ {\it Phys.~Rev.} {\bf #1}}
\newcommand{\PRL}[1]{ {\it Phys.~Rev.~Lett.} {\bf #1}}

\begin{document}
\begin{titlepage}
\setcounter{page}{1}
\renewcommand{\thefootnote}{\fnsymbol{footnote}}

\begin{flushright}
\end{flushright}

\vspace{5mm}
\begin{center}

{\Large\bf Diamagnetism of Confined Dirac Fermions \\ in Disordered Graphene}

\vspace{5mm}

{\bf Ahmed Jellal$^{a,b,c}$\footnote{ajellal@ictp.it, jellal.a@ucd.ac.ma}},
{\bf Malika Bellati}$^{c}$ and {\bf Michael Schreiber}$^{d}$

\vspace{5mm}

$^a${\em Physics Department, College of Science, King Faisal University,\\
PO Box 380, Alahsa 31982,
Saudi Arabia}

$^{b}${\em Max Planck Institute for the Physics of Complex Systems,
N\"othnitzer Str. 38,\\ 01187 Dresden, Germany}

$^{c}${\em Theoretical Physics Group,  
Faculty of Sciences, Choua\"ib Doukkali University},\\
{\em PO Box 20, 24000 El Jadida,
Morocco}

$^{d}${\em Institut f\"ur Physik, Technische Universit\"at, Reichenhainer Str. 70\\
D-09107 Chemnitz, Germany}

\vspace{30mm}

\begin{abstract}

 The diamagnetism of confined Dirac fermions submitted to
 a uniform magnetic field in disordered graphene is investigated.
The solutions of the energy spectrum are used to discuss the orbital magnetism
from a statistical mechanical point of view.
More precisely, by the technique of Green functions
the self-energy for short and long-ranged disorders is obtained. This
allows us to determine the susceptibility for short and long-ranged disorders
together with confinement.
We compare our results with already published work and point out
the relevance of these findings to a systematic formulation of the
diamagnetism in a confining potential.


\end{abstract}
\end{center}
\end{titlepage}



\section{Introduction}

The magnetism of graphene
was first studied as a simple model for three-dimensional
 graphite~\cite{5} where the susceptibility of the disorder-free
graphene was calculated within the effective mass
approximation. It was found that the system exhibits
a large diamagnetism at the Fermi energy $\varepsilon_F = 0$, expressed as a
$\delta$-function of $\varepsilon_F$ at the absolute zero temperature. The
graphene magnetism was considered again in studies on
the graphite intercalation compounds, where the tight-binding
model was applied for a wide range of Fermi
energies~\cite{6,7,8,9}.
The effects of disorder on graphene under magnetic
fields have been examined in early theoretical studies before
the experimental discovery of graphene, where the
electronic structure~\cite{10}, the transport properties~\cite{10,11,12},
and the de Haas-van Alphen effect~\cite{13} were investigated.
Recently the Shubnikov-de Haas oscillation was
studied in disordered graphene~\cite{14,15} and the spectral and
transport properties were examined in presence of lattice
defects under the magnetic fields~\cite{16}.

Very recently, an interesting development on the diamagnetism of disordered graphene
was reported by Koshino and Ando~\cite{Ando1}. They studied
the graphene monolayer orbital magnetism  within the effective mass approximation.
In models of short and long-ranged disorders, the magnetization was calculated with the self-consistent
Born approximation. In the zero-field limit, the susceptibility becomes highly diamagnetic around zero
energy, while it has a long tail proportional to the inverse of the Fermi energy. It was demonstrated how
the magnetic oscillation vanishes and converges to the susceptibility, on going from a strong-field regime
to zero field. Additionally, the behavior at zero energy was shown to be highly singular.

On the other hand,
 an exact solution of a related problem, that has been studied at various levels by researchers
 dealing with different physical issues (see for instance~\cite{villalba,bermudez}),
 was given by one of us (AJ) and his collaborators~\cite{jellal}, considering a relativistic particle subjected to
 an external magnetic field as well as
 to a confining potential. By a similarity transformation 
 the system can be diagonalized in a simple way.
 Solving the eigenvalue equation, and accounting for the complete space of the eigenfunctions one can include various cases related to
 different physical settings. More precisely, from the nature of the problem it was possible to obtain separate angular and radial solutions. The radial
 equation lead to an exact relation between the two-spinor components. In fact, depending on the range of values of three physical quantities,
  the full solution space split into eight disconnected subspaces as summarized in table 1.

Motivated by different investigations
on Dirac fermions in $(2+1)$-dimensions, in particular by references~\cite{Ando1,jellal},
 we treat diamagnetism of a confined system in a statistical
mechanical way.
More precisely, we study the orbital magnetism of Dirac fermions
in uniform magnetic field, disordered graphene and confining potential.
In fact, we combine studies reported in~\cite{Ando1,jellal} to
generalize the results of Koshino and Ando~\cite{Ando1} about diamagnetism
in disordered graphene to the confinement case. This can be done by
using the energy spectrum solutions to study
the self-energy for both regimes: short and long-ranged disorders
by the technique of Green functions. The self-energy allows us to
obtain the shape of the density of states that is needed
to determine the related thermodynamical quantities and discuss
different issues.

Subsequently,
we calculate the susceptibility for two regimes and underline
what makes the difference with respect to the standard case~\cite{Ando1}, namely
the analysis without confinement
 $(\ka=0)$.
For short-ranged disorder, we obtain interesting results
in terms of  a parameter of confinement and disorder strength,  called $C$.
In particular,
we show that there is a quantum correction to the result obtained by Koshino and Ando~\cite{Ando1},
which
disappears  by switching off  $\ka$. Furthermore,
we notice that the clean limit result without confinement can be obtained
by considering $C$ going to zero.

As far as the long-range disorder is concerned, the susceptibility shows  an additional second term of the
order of ${\mathcal O}(C)$ compared to the short-range case. However,  this gives a minor effect since $C$ is assumed
to be small. When the terms of the order ${\mathcal O}(C)^{2}$ are neglected, the susceptibility becomes
just $1-3C$ times as large as in the short-ranged disorder. Accordingly
the integration of  the susceptibility over the energy $\varepsilon$ depends weakly on $C$, while
in the limit $C\rightarrow0$ we again get the susceptibility as $\delta$-function.
Finally, we notice that the case $\ka=0$ allows us to recover the
results and related conclusions proposed in~\cite{Ando1}.

The paper is organized as follows. In section $2$,
we review the energy spectrum solutions of the confinement problem
 needed to deal with different issues.
Section 3 is devoted to introduce disordered graphene where we give
the corresponding four component spinors as well as the disorder
and confining potentials.
In section 4, we use the self-consistent Born approximation to
determine the self-energy and the density of states. These will allow
us to treat the orbital magnetism
by distinguishing between the short- and long-ranged disorders in section 5.
Finally, we conclude by discussing the main results and possible
extension of our work.

\section{Confinement problem}

We start by formulating the problem in terms of our apparoach. In doing so,
we introduce a similarity transformation of the Dirac equation in polar coordinates.
This will be convenient to handle the exact relationship between spinor components and thus derive the
full spectrum accounting for the complete Hilbert space.

\subsection{Hamiltonian}

The problem of a charged particle moving in a constant magnetic field $\vec B =
B\,\hat z$ is a 2D problem in the plane normal to the field [the Cartesian
$(x,y)$-plane or cylindrical $(r,\te)$-plane]. In the relativistic units,
$\hbar = c = 1$, the Dirac equation in $(2+1)$-dimensions for a spinor of
charge $e$ and massless in the electromagnetic potential ${A_\mu } =
({A_0},\vec A)$ reads as follows
\begin{equation}
\left[ {{{i}}{\gamma ^\mu }({\partial _\mu } +
{{i}}e{A_\mu })} \right]\psi  = 0\label{eq1}, \qquad \mu =0,1,2
\end{equation}
where the summation convention over repeated indices is used. ${\gamma ^\mu } =
\left( {{\gamma ^0},\vec \gamma } \right)$ are three unimodular square matrices
satisfying the anti-commutation relation:
\begin{equation}
\left\{ {{\gamma ^\mu
},{\gamma ^\nu }} \right\} = {\gamma ^\mu }{\gamma ^\nu }
 + {\gamma ^\nu }{\gamma ^\mu } = 2{{\cal G}^{\mu
\nu }}
\end{equation}
 where ${\cal G}$ is the metric of
Minkowski space-time, which is equal to $\mbox{diag} ( + \,\, - \,\,- )$. A
minimal irreducible matrix representation that satisfies this relation is given by
${\gamma ^0} = {\sigma _3}$, $  \vec \gamma  = {{i}}\,\vec \sigma$
 where $\left\{ {{\sigma _i}} \right\}_{i= 1}^3$
are the $2\times 2$ hermitian Pauli spin matrices:
\begin{equation}
{\sigma _1} = \left( {\begin{array}{*{20}{c}}
0 & 1  \\
1 & 0  \\
\end{array}} \right),
\qquad
{\sigma _2} = \left( {\begin{array}{*{20}{c}}
0 & -i  \\
i & 0  \\
\end{array}} \right),\qquad
{\sigma _3} = \left( {\begin{array}{*{20}{c}}
1 & 0  \\
0 & -1  \\
\end{array}} \right).
\end{equation}
Equation (\ref{eq1}) can be rewritten as
\begin{equation}
{{i}}{{\partial  \over {\partial t}}}\psi
 = \left( { - {{i}}\,\vec \alpha  \cdot \vec \nabla  + e\vec \alpha
\cdot \vec A } + e A_0\right)\psi \label{eq3}
\end{equation}
where $\vec \alpha $ is the hermitian matrix $\vec \alpha  =
{{i}}\,{\sigma _3}\vec \sigma$. We will see below that
the symmetry of the problem is preserved even if we introduce an additional
coupling to the 2D Dirac-oscillator potential. This coupling is introduced by
the substitution $\vec \nabla
 \to \vec \nabla  + \ka \vec r\si_3$
where $\ka$ is a constant parameter.

For time independent potentials,
the two-component spinor wavefunction $\psi (t,r,\theta )$ is written as
\begin{equation}
\psi (t,r,\theta ) = {e^{- {{i}}\varepsilon t}}\psi (r,\theta )
\end{equation}
and (\ref{eq3})
becomes the energy eigenvalue equation $ ({\cal H} - \varepsilon )\psi  =
0$
 where $\varepsilon$ is the relativistic energy.
The Dirac Hamiltonian ${\cal H}$ is the $2\times 2$ matrix operator
\begin{equation}
{\cal H} =  {{i}}\,{\sigma _3}\vec \sigma  \cdot \hat
r\,{{\cal H}_r} + {{i}}\,{\sigma _3}\vec \sigma  \cdot \hat \theta
\,{{\cal H}_\theta }
\label{eq4}
\end{equation}
where ($\hat r,\hat \theta $) are the unit vectors in cylindrical coordinates
and
\beqn {{\cal H}_r} &=& -i\pa_r + eA_r -i \ka r\sigma_3,
\\
{{\cal H}_{\theta}} &=& -{i\over r} \pa_{\theta} + eA_{\theta}. \nonumber
\label{eq5}
\eeqn
For regular solutions of (\ref{eq3}), square integrability (with respect to the
measure ${d^2}\vec r = r\,dr\,d\theta $) and the boundary conditions require
that $\psi (r,\theta )$ satisfies
\begin{equation}
{\left. {\sqrt r \,\psi (r,\theta )} \right|_{\scriptstyle r = 0}}=0, \qquad
{\left. {\sqrt r \,\psi (r,\theta )} \right|_{\scriptstyle r \to \infty  \hfill}} = 0, \qquad
\psi (\theta+2\pi) = \psi.
\label{eq6}
\end{equation}


To simplify the construction of the solution, we look for a local $2\times2$
similarity transformation $\Lambda (r,\theta )$ that maps the cylindrical
projection of the Pauli matrices $(\vec \sigma  \cdot \hat r$, $\vec \sigma
\cdot \hat \theta )$ into their canonical Cartesian representation $({\sigma
_1}, {\sigma _2})$, respectively~\cite{explain}. That means
\begin{equation}
\Lambda\, \vec \sigma  \cdot \hat r\,{\Lambda ^{ - 1}} = {\sigma _1},
\qquad
\Lambda \,\vec \sigma  \cdot \hat \theta\,{\Lambda ^{ - 1}} = {\sigma _2}.
\label{eq7}
\end{equation}
A $2\times 2$ matrix that satisfies this requirement is
\begin{equation}
\Lambda (r,\theta ) = \lambda (r,\theta )\,{e^{{{}{{{i}}
\over 2}}{\sigma _3}\theta }} 
\label{eq8}
\end{equation}
where $\lambda (r,\theta )$ is a real function and the exponential
is a $2\times 2$ unitary matrix. The Dirac Hamiltonian (\ref{eq4})
gets mapped into
\begin{equation}
H = \Lambda {\cal H}{\Lambda ^{ - 1}} =  -
{\sigma _2}{H_r} + {\sigma _1}{H_\theta }\label{eq9}
\end{equation}
where
\beqn
H_r &=&-i\left(\partial _r - {\lambda _r \over\lambda }\right) + i e{A_r} - i\ka r\sigma_3, \\
H_{\theta} &=& - {i \over r}\left(\partial _\theta  - {\lambda _\theta \over
\lambda}  -{i\over 2}\sigma_3 \right)+ eA_\theta\nonumber \label{eq10}
\eeqn
with  ${\lambda _k} = {\partial _k}\lambda $. Therefore, the 2$\times{}$2
Dirac Hamiltonian becomes
\begin{equation}
\label{smham}
H =
\small
\begin{pmatrix}
 0  & \partial _r - {\lambda _r \over\lambda } +
{1 \over 2r} + i e A_r - \ka r - {i \over r}\left( \partial_\theta  -
{\lambda _\theta  \over \lambda } \right) + e A_\theta \\
-\partial _r + {\lambda _r\over \lambda } -
{1 \over 2r} - i e A_r - \ka r - {i \over r} \left(\partial _\theta  -
{\lambda _\theta \over \lambda } \right) + e A_\theta
   & 0
\end{pmatrix}.
\end{equation}
Thus, hermiticity of (\ref{smham}) requires that
\begin{equation}
{\lambda _\theta } = 0, \qquad
{{{\lambda _r}} \over
\lambda } - {{1 \over {2r}}}=0
\label{eq12}
\end{equation}
and fixes the exact form of the modulus of the similarity transformation to be
$\lambda (r,\theta ) = \sqrt r$. It is interesting to note that ${\lambda ^2}$
turns out to be the integration measure in 2D cylindrical coordinates. We could
have eliminated the $\lambda$ factor in the definition of $\Lambda$ in
(\ref{eq8})
by proposing that the new spinor wavefunction {$\chi$} be replaced by ${{1
\over {\sqrt r }}}\chi (r,\theta )$. In that case, the transformation matrix
$\Lambda{}$ becomes simply ${e^{{{}{{{i}} \over 2}}{\sigma _3}\theta }}$, which
is unitary. However, making the presentation as above gave us a good
opportunity to show (in a different approach) why it is customary to take the
radial component of the wavefunction in 2D cylindrical coordinates to be
proportional to ${1 \over \sqrt r }$. Finally, we obtain the
$(2+1)$-dimensional Dirac equation $ \left( {H - \varepsilon } \right)\chi  =
0$ for a charged spinor in static electromagnetic potential as
\begin{equation}
\left( {\begin{array}{*{20}{c}}
{- \varepsilon } & {{\partial _r} + {{i}}e{A_r} -
\ka r - {{{{i}} \over r}}{\partial _\theta } +
e{A_\theta }}  \\
{ - {\partial _r} - {{i}}e{A_r} - \ka r -
{{}{{{i}} \over r}}{\partial _\theta } + e{A_\theta }} & { - \varepsilon }  \\
\end{array}} \right)\left( {\begin{array}{*{20}{c}}
{\mathop {{\chi _ + }(r,\theta )}\limits_{} }  \\
{{\chi _ - }(r,\theta )}  \\
\end{array}} \right) = 0\label{eq13}
\end{equation}
where ${\chi_\pm }$ are the components of the transformed wavefunction $\left|
\chi  \right\rangle  = \Lambda \left| \psi  \right\rangle $. This equation will
be solved by choosing an appropriate gauge to end up with the full Hilbert
space.

\subsection{Energy spectrum}

Now, we specialize to the case where a constant magnetic field of strength $B$
is applied at right angles to the ($r, \theta $)-plane, which is $\vec B =
B\,\hat z$. Therefore, the electromagnetic potential has the time and space
components
\begin{equation}
{A_0} = 0, \qquad \vec A(r,\theta ) =
{{1 \over 2}}Br\,\hat \theta.
\end{equation}
Consequently, (\ref{eq13})
becomes completely separable and we can write the spinor wavefunction as
\begin{equation}
{\chi _ \pm }(r,\theta ) = {\phi _ \pm
}(r)\,\tau (\theta ).
\end{equation}
Thus, the angular component satisfies $ - {{i}}{{}{{d\tau } \over {d\theta }}}
= \xi \,\tau $ where {$\xi{}$} is a real separation constant giving the
function
\begin{equation}
\tau (\theta ) = {{1
\over {\sqrt {2\pi } }}}{e^{{{i}}\xi \theta }}.
\end{equation}
On the other hand, the boundary condition $\psi (\theta  + 2\pi ) = \psi
(\theta)$ requires that
${e^{{{i}}\,2\pi \xi }}{e^{ - {{i}}{\sigma _3}\pi }} =  + 1$ which, in turn,
demands that $ {e^{{{i}}\,2\pi \xi }} =  - 1 $ giving the following quantum
number:
\begin{equation}
\xi  = {{1 \over 2}} v, \qquad
v =\pm 1, \pm 3, \pm 5 \cdots.
\label{eq14}
\end{equation}
Consequently, the Dirac equation for the two-component radial spinor is reduced
to
\begin{equation}\left( {\begin{array}{*{20}{c}}
{- \varepsilon } & {{{d \over {dr}}} +
{{\xi  \over {r}}} + \om r}  \\
{ - {{d \over {dr}}} + {{\xi  \over {r}}} +
\om r} & { - \varepsilon }  \\
\end{array}} \right)\left( {\begin{array}{*{20}{c}}
{\mathop {{\phi _ + }(r)}\limits_{} }  \\
{{\phi _ - }(r)}  \\
\end{array}} \right) = 0\label{eq15}
\end{equation}
where the physical constant $\om$ is given by $ \om = l_B^{-2}  - \ka$ and $l_B$
is the magnetic length $l_B=\frac{1}{\sqrt{eB}}$. Thus, the presence of the
2D Dirac-oscillator coupling did, in fact, maintain the symmetry of the problem
as stated below (\ref{eq3}). Moreover, its introduction is equivalent to
changing the magnetic field as $eB \lga eB - 2\ka$.
 As a result of
the wave equation (\ref{eq15}), the two spinor components satisfy the "kinetic
balance" relation
\begin{equation}
{\phi _ \mp }(r) = \frac{1}{{\varepsilon}}\left[ { \mp
\frac{d}{{dr}} + \frac{\xi} {r} + \om r} \right]{\phi _ \pm }(r)
\label{eq16}
\end{equation}
where $\varepsilon  \ne  0$. Therefore, the solution of the problem with the
top/bottom sign corresponds to the positive/negative energy solution. Using the
exact relation (\ref{eq16})
to eliminate one component in terms of the other in (\ref{eq15}) results in the
following Schr\"{o}dinger-like differential equation for each spinor component:
\begin{equation}
\left\{ { - \frac{{{d^2}}}{{d{r^2}}} + \frac{{{{}{\xi}}\left( {{{}{\xi}} \mp 1}
\right)}}{{{r^2}}} + {\om^2}{r^2} + \left[ {- {\varepsilon ^2} +
\om\left( {2\xi  \pm 1} \right)} \right]} \right\}{\phi _ \pm }(r) = 0.
\label{eq17}
\end{equation}
We stress that this equation gives {only one} radial spinor
component. One must choose either the top or bottom sign to obtain the
component that corresponds to the positive or negative energy solutions,
respectively. The second component is obtained by substituting this into the
exact relation (\ref{eq16}). Nonetheless, we only need to find one
solution (the positive- or negative-energy solution), because the other is
obtained by a simple map. For example, the following map takes the positive
energy solution into the negative energy solution:
\begin{equation}
\varepsilon  \lga  - \varepsilon,\qquad v  \lga  - v,\qquad \om \lga
- \om,\qquad {\phi _ \pm } \lga {\phi _ \mp }
\label{eq18}
\end{equation}
which, in fact, is the $\cal{CPT}$ transformation. Here the charge conjugation
$C$ means that $e \lga  - e$ and  $\ka\lga-\ka$. 
It is easy to check that the above map (\ref{eq18}) originates
from the fact that the Dirac equation (\ref{eq15}) is invariant under such
transformation. Hence, we just need to solve for positive energies and use the
above transformation to obtain the negative energy solutions. The total spinor
wavefunction reads
\begin{equation}
\psi (r,\theta ) = \frac{1}{{\sqrt r }}\,{e^{{{}{{{i}}
}}\xi \theta }} {e^{ - {{}{{{i}} \over 2}}{\sigma
_3}\theta }} \phi (r)
\label{eq19}
\end{equation}
where $\phi (r)$ has two components, such as
\begin{equation}
\phi= \left( {\begin{array}{*{20}{c}}
 \phi _ + \\
 \phi _ -\\
\end{array}} \right).
\end{equation}
Equation (\ref{eq17}) looks like the non-relativistic oscillator problem with a
certain parameter map of the frequency, angular momentum, and energy. For
regular solutions of (\ref{eq17}),
the bound states 
will be of the form
\begin{equation}
{\phi _ \pm }\sim
 {z^\mu }{e^{ - {z \mathord{\left/
{\vphantom {z 2}} \right.
\kern-\nulldelimiterspace} 2}}}L_n^\nu (z)
\label{eq20}
\end{equation}
where $L_n^\nu (z)$ is the associated Laguerre polynomials of order $n =
0,1,2,\cdots$ and $z = {\rho ^2}{r^2}$. The constants $\left\{ {\mu ,\nu ,\rho
} \right\}$ are real and related to the physical parameters $B$, {$\ka$}, and
{$\xi$}. Square integrability and the boundary conditions require that $2\mu
\ge {{}{1 \over 2}}$ and $\nu  >  - 1$.

Substituting the ansatz (\ref{eq20})
into (\ref{eq17})
and using the differential equation for the Laguerre polynomials~\cite{grad},
we obtain four equations. Three of them determine the parameters
$\left\{ {\mu ,\nu ,\rho } \right\}$ and one determines the energy spectrum.
The first three are \beqn 2\mu &=& \nu + {{1 \over 2}}, \qquad
\rho^2=|\om|, \nonumber\\
\nu  &=& \pm \left\{ {\begin{array}{*{20}{c}}
\xi -{1\over 2 }, \qquad  { \varepsilon > 0}   \\
\xi +{1\over 2}, \qquad
{\varepsilon < 0}.  \\
\end{array}}
\right. \label{eq21}
\eeqn
For regular solutions of (\ref{eq17}), the $\pm $ sign in the expression
for $\nu $ corresponds to $\pm \xi > 0$. The fourth equation gives the
following (positive and negative) energy spectra:
\begin{equation}\label{2.22}
\varepsilon _{n,\xi}^{\pm}  = \pm \sqrt{{2|\om|}
\left[2n+1 \pm {s-s'\over 2} +\xi (s+s')\right]}
\end{equation}
where $s=\sf{sgn}\left (\om \right)={|\om|\over \om}$ and $s'=\sf{sgn}\left
(\xi \right)$. The sign of $\om$ depends on whether
$\ka $ is larger or smaller than the magnetic length {$l$}$_{B}$.
To compare our work with frequently used notation in the literature, we can
replace the quantum number {$\xi$} by $k + {1\over 2}$, where $k = 0, \pm 1,
\pm 2,\cdots $ and $\xi \lga -\xi$ implies that $k \lga -k-1$. In that case, one
may write the positive eigenvalues as
\begin{equation}\label{2.22a}
\varepsilon _{n,k}^{+}  = \sqrt{ {2|\om|}
\left[2n +1 +s +{k} (s+s')\right]}
\end{equation}
and the negative ones as
\begin{equation}\label{2.22b}
\varepsilon _{n,k}^{-}  = - \sqrt{ {2|\om|}
\left[2n +1+s' +{k} (s+s')\right]}
\end{equation}
where $s'= +1$ for $k$ = 0. It is interesting to note that for $\xi \om  < 0$ the
spectrum is infinitely degenerate because it is independent of $\xi$. However,
for $ \xi \om > 0$ the degeneracy is finite and equal to $n + k + 1$.
Substituting the wavefunction parameters given by (\ref{eq21}) into the ansatz
(\ref{eq20}) gives for $\varepsilon  > 0$:
\begin{equation}
{\phi _ + }(r) =  {x^{\left| k +{1\over 2}  \right|}} \, e^{ - {1 \over 2}x^2}
\\
 \left\{ {\begin{array}{*{20}{c}}
A_{n,k}^{++}\, L_n^{k}\left(x^2\right),\qquad~~~ &  {k  \geq 0}  \\
A_{n,k}^{+-} \, x
L_n^{-k} \left(x^2\right), \qquad & {k  < 0}  \\
\end{array}} \right.
\label{eq22}
\end{equation}
as well as for $\varepsilon  < 0$:
\begin{equation}
{\phi _ - }(r) =  {x^{\left| k +{1\over 2}  \right|}} \, e^{ - {1 \over 2}x^2}
\\
 \left\{ {\begin{array}{*{20}{c}}
A_{n,k}^{-+}\, x L_n^{k+1}\left(x^2\right),\qquad &  {k  \geq 0}  \\
A_{n,k}^{--} \,
L_n^{-k-1} \left(x^2\right), \qquad & {k  < 0}  \\
\end{array}} \right.
\label{eq23}
\end{equation}
where $x=r\sqrt{|\om|}$ and ${A_{n,k}^{ij} }$ are normalization constants that
depend on the physical quantities {$l_B $} and {$\ka$. The lower
component ${\phi _ - }(r)$ is obtained by substituting (\ref{eq22}) and (\ref{eq23})  into the
exact relation (\ref{eq16}). Doing so while exploiting the
differential and recursion properties of the Laguerre polynomials
 we obtain the following for $\varepsilon  > 0$:
\beqn
\label{2.26'} {\phi_-}(r) &=&  \frac{{\sqrt{\left|\om \right|}}}
{{\varepsilon^+_{n,k}}}\,
x^{\left|k+{1\over 2} \right|}\,  e^{ -{1 \over 2}x^2}\nonumber\\
 &&\times \left\{ {\begin{array}{*{20}{ccc}}
{A_{n,k}^{++}}\, x {\left[(s-1) L_n^{k}\left(x^2\right) +
2L_n^{k+1}\left(x^2\right)\right]},
~~~~~~~~~~~~~~~~~~~~~~~~~~~~~~~~ &  {k \geq 0}  \\
{A_{n,k}^{+-}}\; \left[(s-1) (n-k) L_n^{-k-1}\left(x^2\right)-(s+1) (n+1)
L_{n+1}^{-k-1}\left(x^2\right)\right],  &
{k  < 0}.  \\
\end{array}} \right.
\eeqn
On the other hand, repeating the same calculation for the upper component of
the negative energy solution gives the function \beqn\label{2.27'} {\phi_+}(r)
&=& \frac{{\sqrt{\left|\om \right|}}} {{\varepsilon^-_{n,k}}}\,
x^{\left|k+{1\over 2}\right|}\;  e^{ -{1 \over 2}x^2}\nonumber\\
&& \times
 \left\{ {\begin{array}{*{20}{ccc}}
{A_{n,k}^{-+} }\, \left[(1+s) (n+k+1) L_n^{k}\left(x^2\right) +(1-s) (n+1) L_{n+1}^{k}\left(x^2\right)\right],  & {k  \geq 0}  \\
A_{n,k}^{--}\, x {\left[(1+s) L_n^{-k-1}\left(x^2\right) -
2L_n^{-k}\left(x^2\right)\right]},
 ~~~~~~~~~~~~~~~~~~~~~~~~~~~&  {k  < 0}  \\
\end{array}} \right.
\eeqn
which can have also been obtained by applying the $\cal{CPT}$ map
(\ref{eq18}) to (\ref{2.26'}). Thus, the structure of the whole Hilbert space
solution consists of eight disconnected
spaces that can be displayed in tabular form as shown in Table 1:\\
\begin{center}
\begin{tabular}{|c|c|c|c|c|c|c|c|}
\hline
{\bf Frequency} & $\ka  > l_B^{-2}$      & $\ka  < l_B^{-2} $   \\[4mm]
\hline
{\bf Energy} & $\varepsilon  > 0$  \vline ~ $\varepsilon  < 0$ &
$\varepsilon  > 0$ \vline ~  $\varepsilon  < 0$\ \\[4mm] \hline
{\bf Azimuth} & $ k\geq  0$  \vline ~ $k <  0$  \vline ~  $ k\geq  0$  \vline ~ $k <  0$   & $ k\geq  0$ \vline ~ $k <  0$
\vline ~ $ k\geq  0$ \vline ~  $k <  0$        \\[4mm] \hline
\end{tabular}\\
\end{center}
\vspace{1mm}
\begin{center}
{Table 1: Complete space solution.}
\end{center}
Using the standard definition, we calculate all  normalization
constants in the above wavefunctions. These are summarized
in the Table 2, where as stated above $s={\sf{sgn}}(\om)=|\om|/\om$:\\

\begin{center}
\begin{tabular}{|c|c|c|}
\hline
\bf  Energy   &\bf Azimuth &\bf  Normalization\\[4mm] \hline
{$\varepsilon  > 0$} & $k \geq 0$ & ~~ $A_{n,k}^{++} = \sqrt{\frac{2n!} { \pi (n+k)!}
\left[{4{|\om|}\over (\varepsilon^+_{n,k})^2} \left\{ 2(n+k+1) +
 n({{1-s}})\right\}\right]^{-1}}$\ \\[6mm] \hline
 $\varepsilon  > 0$
&  $k <  0$ & ~~~ $~~A_{n,k}^{+-} = \sqrt{\frac {n!}{\pi (n-k)!}
\left[{2{|\om|}\over (\varepsilon^+_{n,k})^2}
\left\{2(n+1) + (k+1) (s-1)\right\}\right]^{-1}}$ \ \\[6mm] \hline
$\varepsilon <  0$ &     $k \geq 0$ &\   $A_{n,k}^{-+} = \sqrt{\frac{n!}{ \pi (n+k+1)!}
\left[{2{|\om|}\over (\varepsilon^-_{n,k})^2} \left\{ 2(n+1) + k(s+1)
\right\}\right]^{-1}}$           \\[6mm] \hline
 $\varepsilon <  0$
&  $k <  0$  &~~~       $A_{n,k}^{--} = \sqrt{\frac{2n!}{ \pi (n-k-1)!}
\left[{4{|\om|}\over (\varepsilon^-_{n,k})^2} \left\{ 2(n-k) + n(s+1)\right\}\right]^{-1}}$~~~
\\[6mm] \hline
\end{tabular}
\end{center}
\vspace{1mm}
\begin{center}
 {Table 2: Normalization in terms of different physical quantities.}
\end{center}

\section{Disordered sublattices}

Our main task is to analyze the diamagnetism
of confined Dirac fermions in disordered graphene. In the previous section,
we settled the required tools for the confinement, however
we still need to introduce the disorder potential
and related matter. For this, we write the eigenspinors
as well as the impurity and confining potentials in four components.


Having obtained the energy spectrum solutions corresponding to one sublattice, it
is worthwhile to deal with a system of graphene.
Such a system
 is composed of a honeycomb network of carbon atoms where a unit
cell contains  one atom each from
a pair of sublattices, denoted by ${\rm I}$ and ${\rm II}$.
It can be described by a $4\times 4$ matrix Hamiltonian, such as
\begin{equation}\label{hamilfour}
    H=\left(
        \begin{array}{cccc}
          0 & a & 0 & 0 \\
          a^{\da} & 0 & 0 & 0 \\
          0 & 0 & 0 & a^{\da} \\
          0 & 0 & a & 0 \\
        \end{array}
      \right)
\end{equation}
where in polar coordinates the two operators $a$ and $a^\da$ take the form
\begin{eqnarray}
  a &=& \partial _r - {\lambda _r \over\lambda } +
{1 \over 2r} + i e A_r - \ka r - {i \over r}\left( \partial_\theta  -
{\lambda _\theta  \over \lambda } \right) + e A_\theta \\
  a^{\da} &=& -\partial _r + {\lambda _r\over \lambda } -
{1 \over 2r} - i e A_r - \ka r - {i \over r} \left(\partial _\theta  -
{\lambda _\theta \over \lambda } \right) + e A_\theta.
\end{eqnarray}
Clearly, they act on  one component of the eigenspinors
$F_{\rm I}^{K},F_{\rm II}^{K},F_{\rm I}^{K^{\prime}},F_{\rm II}^{K^{\prime}}$ where
$F_{\rm I}^{K},F_{\rm II}^{K}$ represent the envelope functions at ${\rm I}$ and ${\rm II}$
sites for  point $K$, respectively, and
$F_{\rm I}^{K^{\prime}},F_{\rm II}^{K^{\prime}}$ for $K^{\prime}$, with $K$
and $K'$ indexing the two valleys.

It is convenient for our task to label the eigenspinors  by $\al=(j,n,k)$ with the valley index
$j=K,K^{\prime}$, the Landau level index $n=0,\pm1,\cdots$ and the wavevector
$k$. The eigenspinors for $k\geq0$ and $\varepsilon>0$
can easily be deduced from the above analysis. For $K$ we have
\begin{equation}\label{wavefunK1}
     F^{K}_{n,k} (r,\theta)= {A_{n,k}^{++}}\ {\left|\om \right|}^{1\over 4}\,{e^{{{}{{{i}}
}}(k+\frac{1}{2}) \theta }} {e^{ - {{}{{{i}} \over 2}}{\sigma
_3}\theta }} x^{{\left|k +{1\over 2}\right|}-{1\over 2}} \, e^{ - {1 \over 2}x^2} 
\\
\left(
  \begin{array}{c}
    L_n^{k}\left(x^2\right) \\
\frac{{\sqrt{\left|\om \right|}}}
{{\varepsilon^+_{n,k}}} \, x {\left[(s-1) L_n^{k}\left(x^2\right) +
2L_n^{k+1}\left(x^2\right)\right]}\\
    0 \\
    0 \\
  \end{array}
\right)
\end{equation}
as well as
\begin{equation}\label{wavefunK2}
    F^{K^{\prime}}_{n,k}(r,\theta) = {A_{n,k}^{++}}\ {\left|\om \right|}^{1\over 4}\,{e^{{{}{{{i}}
}}(k+{1\over 2}) \theta }} {e^{ - {{}{{{i}} \over 2}}{\sigma
_3}\theta }} {x^{\left| k +{1\over 2}  \right|-{1\over 2}}} \, e^{ - {1 \over 2}x^2} 
\\\left(
    \begin{array}{c}
      0 \\
      0 \\
      \frac{{\sqrt{\left|\om \right|}}}
{{\varepsilon^+_{n,k}}} \, x {\left[(s-1) L_n^{k}\left(x^2\right) +
2L_n^{k+1}\left(x^2\right)\right]} \\
      L_n^{k}\left(x^2\right) \\
    \end{array}
  \right)
\end{equation}
 for the point $K'$. Without loss of generality in the forthcoming analysis
  we only focus on the case where
$k\geq0$ and $\varepsilon>0$.
Other configurations can be recovered from the first one
by making use of some mapping.

To consider disordered graphene and make comparisons with already published
work~\cite{Ando1}, we introduce
the disorder potential characterized by two simple models:
short- and long-ranged scatterers~\cite{10}.
The first is an on-site potential
localized at a particular ${\rm I}$ or ${\rm II}$ site with a random amplitude.
A scatterer on  site ${\rm I}$ at $\Vec{R}_{\rm I}$
is represented as
\begin{equation}
 U(\Vec{r}) =
\left(
\begin{array}{cccc}
 1 & 0 & z_{\rm I}^* z'_{\rm I} & 0 \\
 0 & 0 & 0 & 0 \\
z_{\rm I} {z'_{\rm I}}^* & 0 & 1 & 0 \\
 0 & 0 & 0 & 0
\end{array}
\right)
u_i \delta(\Vec{r}-\Vec{R}_{\rm I})
\end{equation}
and that on site ${\rm II}$ at $\Vec{R}_{\rm II}$ as
\begin{equation}
 U(\Vec{r}) =
\left(
\begin{array}{cccc}
 0 & 0 & 0 & 0 \\
0 & 1 & 0& z_{\rm II}^* z'_{\rm II} \\
 0 & 0 & 0 & 0 \\
0 & z_{\rm II} {z'_{\rm II}}^*  & 0 & 1
\end{array}
\right)
u_i  \delta(\Vec{r}-\Vec{R}_{\rm II})
\end{equation}
where we introduced
$z_{\rm X} = e^{i\Vec{K}\cdot\Vec{R}_{\rm X}}$, $z'_{\rm X} = e^{i\Vec{K}'\cdot\Vec{R}_{\rm X}}$
with ${\rm X = I}$ and ${\rm II}$, and $u_i = (\sqrt{3}a^2/2)U_i$
with the on-site energy $U_i$.
We assume that the scatterers are equally distributed
on ${\rm I}$ and ${\rm II}$ sites with density $n_i^{\rm I} = n_i^{\rm II} = n_i/2$ and the
mean square amplitude $\av{(u_i^{\rm I})^2}=\av{(u_i^{\rm II})^2}=u_i^2$.

Dominant scatterers in graphene are expected to have a potential
range larger than the lattice constant for which inter-valley scattering
is much smaller than intra-valley scattering.  Further, realistic
scatterers are likely to have the range comparable to the Fermi
wavelength~\cite{Nomu06,Ando06,Nomu07}. In the following, however, we shall
assume scatterers with potential range smaller than the Fermi
wavelength.  The reason is that the results are expected~\cite{Ando1} to remain
qualitatively the same and further that actual calculations are
practically possible.

In this long-range model, a scatterer at $\Vec{R}$ is expressed by
\begin{equation}
 U(\Vec{r}) =
\left(
\begin{array}{cccc}
 1 & 0 & 0 & 0 \\
0 & 1 & 0& 0 \\
 0 & 0 & 1 & 0 \\
0 &  0  & 0 & 1
\end{array}
\right)
u_i  \delta(\Vec{r}-\Vec{R}).
\end{equation}
We assume the scatterer density $n_i$ and the mean square amplitude $u_i^2$.
It was shown that the transport properties in the short-ranged disorder
and the long-ranged one are qualitatively similar~\cite{10,11,12}. 

To complete our model
we consider the confining potential as $4\times 4$ matrix
as well. This is
\begin{equation}\label{confpoten}
    V(r)=\left(
           \begin{array}{cccc}
             \kappa r & 0 & 0 & 0 \\
             0 & -\kappa r & 0 & 0 \\
             0 & 0 & \kappa r & 0 \\
             0 & 0 & 0 & -\kappa r \\
           \end{array}
         \right)
\end{equation}
which of course can easily be obtained from the former study.
The above established mathematical tools will serve us to
deal with our task. More precisely, we will see
how it can be used to describe the diamagnetism of
relativistic particles under the influence of
three constraints. These are: magnetic field,
confined and disordered potentials.

\section{Self-consistent Born approximation}

As we claimed before, one of our objectives
is to generalize the results obtained in
\cite{Ando1} to the confinement case.
Actually, this can be achieved, for instance, by adopting the same
method as~\cite{Ando1}, based
on application of  the self-consistent Born approximation
to evaluate the self energy and therefore the density of states.
This allows us to determine
some thermodynamical quantities, in particular the
susceptibility.

\subsection{Self-energy}

To deal with different issues,
 we introduce the Green function
that is related to the  self-energy $\Sigma$ via the Dyson
equation. It is
\beq
\langle G_{\alpha\alpha'}(\varepsilon)\rangle=
\delta_{\alpha\alpha^{\prime}}G_{\alpha}^0(\varepsilon)+
G_{\alpha}^0(\varepsilon)
\sum_{\alpha''} \Sigma_{\alpha\alpha''}(\varepsilon)
\langle G_{\alpha''\alpha'}(\varepsilon)\rangle
\eeq
where the first term contains the matrix elements of the unperturbed Green function corresponding to the Hamiltonian
considered before, such as
\beq
G_{\alpha\alpha'}^0(\varepsilon)=
\langle \al|\frac{1} {\varepsilon- H} |\al'\rangle=
\delta_{\alpha\alpha^{\prime}}G_{\alpha}^0(\varepsilon).
\eeq
To  proceed further, we make use of
the self--consistent Born approximation for our system.
The self-energy of the disorder-averaged Green function $\langle
G_{\alpha,\alpha^{\prime}}\rangle$ can be written as
\begin{equation}  \label{selfe}
\Sigma_{\alpha\alpha^{\prime}}(\varepsilon)=\sum_{\alpha_{1}\alpha_{2}} \langle
U_{\alpha\alpha_{1}}U_{\alpha_{2}\alpha^{\prime}}\rangle\langle
G_{\alpha_{1}\alpha_2}(\varepsilon)\rangle
\end{equation}
where the symbol $\langle\cdots\rangle$ represents the average
over the impurity configurations.

The above equation can be solved by considering the range of the disorder.
Specifically, in the short-ranged model one can show that
the self-energy and the averaged Green function are diagonal with respect to $\alpha$~\cite{Ando1}.
Furthermore,  the self-energy becomes
independent of $\alpha$ and leads to
\begin{equation}  \label{self}
\langle
G_{\alpha\alpha^{\prime}}(\varepsilon)\rangle=\delta_{\alpha\alpha^{%
\prime}}G_{\alpha}(\varepsilon)
\end{equation}
where  $G_{\alpha}(\varepsilon)$ is given by
\begin{equation}  \label{fungre}
G_{\alpha}(\varepsilon)=G(\varepsilon,\varepsilon_{\alpha})\equiv\frac{1}{%
\varepsilon-\varepsilon_{\alpha}-\Sigma(\varepsilon)}.
\end{equation}
From (\ref{selfe}) and (\ref{fungre}), it is clear that
one has to find the appropriate solution for $\Sigma$. To do so,
one needs to introduce respective approximations.

Having described the needed tools, let us see how they can be applied
to analyze the basic features of the present system.
According the former analysis we have
two contributions to the total self-energy $\Sigma(\varepsilon)\equiv\Sigma_{\sf{tot}}(\varepsilon)$,
namely
\begin{equation}  \label{twocontr}
\Sigma_{\sf{tot}}(\varepsilon)=\Sigma_{\sf{%
dis}}(\varepsilon)+\Sigma_{\sf{conf}}(\varepsilon)
\end{equation}
where $\Sigma_{\sf{
dis}}$ and $\Sigma_{\sf{conf}}$  correspond to the disordered
and confining potentials, respectively. In the forthcoming analysis,
we separately determine each part. Note that, what makes a difference
with respect to the study reported in~\cite{Ando1} is
the second contribution and therefore one can see its impact
on such study.

As a first step, we have to evaluate the
 matrix elements of different potentials. For the
 impurity potential, using the eigenspinors it is straightforward to show
\begin{equation}  \label{shortrang}
\langle\alpha|U_{i}|\alpha'%
\rangle=\sqrt{|\om|}u_{i}
\delta_{\alpha\alpha^{\prime}}
\end{equation}
where
$\om$ is the parameter introduced in (\ref{eq15})
and the integration is performed over the coordinates $r$ and $\theta$.
This can be used together with (\ref{selfe})
for the short-ranged potential, to straightforwardly obtain
\begin{equation}  \label{selfenergy}
\Sigma_{\sf{dis}}(\varepsilon)=\frac{n_{i}u_{i}^{2}}{2}%
|\om|\sum^{\infty}_{n=-\infty}\sum_{k=0}^{\infty}\frac{g(\varepsilon_{n,k})} {%
\varepsilon-\varepsilon_{n,k}-\Sigma_{\sf{tot}}(\varepsilon)}
\end{equation}
where the cutoff function $g(\varepsilon)$ is given
\begin{eqnarray}
 g(\varepsilon)=\left\{%
\begin{array}{ll}
    1, & \qquad |\varepsilon|< \varepsilon_c \\
    0, & \qquad  \mbox{otherwise}. 
\end{array}%
\right.
\end{eqnarray}

As far as the confining potential is concerned,  the corresponding
 matrix elements can be evaluated to end up with
\begin{eqnarray}  \label{matconfpo}
\langle\alpha |\pm\kappa r|\alpha^{\prime}\rangle
&=& \sqrt{|\om|}\ \kappa
\Bigg[1-\frac{|\om|}{\varepsilon_{n^{%
\prime},k^{\prime}}^{+} \varepsilon_{n,k}^{+}}2s(n+k+1) \Bigg]\
\delta_{
\alpha\alpha^{\prime}}.
\end{eqnarray}
This leads to the self-energy for the confinement:
\begin{equation}  \label{selfconfpo}
\Sigma_{\sf{conf}}(\varepsilon) = \frac{\kappa^{2}}{2}%
|\om|\sum^{\infty}_{n=-\infty}\sum_{k=0}^{\infty} \left[1-\frac{s(n+k+1)} {%
2n+(s+1)(k+1)} \right]^{2}\frac{g(\varepsilon_{n,k})} {%
\varepsilon-\varepsilon_{n,k}-\Sigma_{\sf{tot}}(\varepsilon)}.
\end{equation}
It is clear that $\Sigma_{\sf{conf}}$ is strongly $\kappa$-dependent, which
is an expected result because of the confining potential expression (\ref{confpoten}).

So far, we have obtained the different contributions to
 the total self-energy. This can be written as
\begin{eqnarray}
  \Sigma_{\sf{tot}}(\varepsilon) = \frac{|\om|}{2} \sum^{\infty}_{n=-%
\infty}\sum_{k=0}^{\infty}
\left\{{n_{i}u_{i}^{2}}+\kappa^{2}\left[1-\frac{%
s(n+k+1)}{2n+(s+1)(k+1)} \right]^{2}\right\} \frac{g(\varepsilon_{n,k})} {\varepsilon-\varepsilon_{n,k}-%
\Sigma_{\sf{tot}}(\varepsilon)}.\label{toexa}
\end{eqnarray}
This expression can be simplified by
choosing $s=1$ and requiring the condition
$n+k+1\neq 0$. In this case, (\ref{toexa}) reduces to
\begin{eqnarray}\label{toexa2}
  \Sigma_{\sf{tot}}(\varepsilon) = \frac{C|\om|}{2} \sum^{\infty}_{n=-%
\infty}\sum_{k=0}^{\infty}
 \frac{g(\varepsilon_{n,k})} {\varepsilon-\varepsilon_{n,k}-%
\Sigma_{\sf{tot}}(\varepsilon)}.
\end{eqnarray}
where $C$ is a dimensionless parameter defined as
\begin{equation}  \label{paramc}
C=n_{i} u_{i}^{2}+\frac {\ka^{2}} {4}
\end{equation}
which  depends on the set of parameters. Thus, one can fix them
to derive specific results and offer different interpretations.

At this stage, one can inspect the above results to underline their
basic properties. One way to do so is to look at the case $\ka =0$, i.e.
without confining the system.
This simply  reduces  $C$ to
the so-called disorder strength
\begin{equation}\label{const}
    C|_{\ka=0}=n_{i}u_{i}^{2}
\end{equation}
which leads to the self-energy
\begin{equation}\label{self}
\Sigma(\varepsilon)=\frac{n_{i}u_{i}^{2}}{2}l_{B}^{-2}\sum_{n=-\infty}^{\infty}
\frac{g(\varepsilon_{n})}{\varepsilon-\varepsilon_{n}-\Sigma(\varepsilon)}
\end{equation}
where the corresponding eigenvalues are
\begin{equation}\label{eneg}
    \varepsilon_{n}=l_B^{-2}{\sf{sgn}}(n)\sqrt{|n|}.
\end{equation}
This was obtained by studying the Dirac fermions in magnetic field
and disordered graphene, more details can be found in~\cite{Ando1}. On the other hand,
comparing (\ref{paramc}) and (\ref{const}),
$C$ can be interpreted as a parameter of confinement and disorder strength.

\subsection{Density of states}

As we claimed before,
the density of states is strongly needed
and will play a crucial role in the forthcoming
analysis. Specifically, it is related to different thermodynamical
quantities and therefore allows us to determine them in an appropriate
way. This statement will be clarified starting from next section.

For later convenience, we consider
the density of states used in~\cite{16} by dealing with
some features of graphene. This is
\begin{equation}  \label{densitysta1}
\rho(\varepsilon)=-\frac{1}{\pi}\sum_{\alpha}{\sf{Im}}G_{\alpha}(%
\varepsilon+i0).
\end{equation}
This form can be handled by fixing different conditions. For this,
we distinguish between short- and long-ranged disorders.
Returning to our results,  we have
\begin{equation}  \label{densitysta2}
\rho(\varepsilon)=-\frac{1}{\pi}\frac{2}{C|\om|}\ {\sf{Im}}\Sigma_{\sf{%
tot}}(\varepsilon+i0)
\end{equation}
for short-ranged disorder.

Let us recall that
the transport properties in the short- and long-ranged disorders
are qualitatively similar~\cite{Ando1}. In the last case, the self-energy and
Green function have off-diagonal matrix elements between $(j,n,k)$ and $%
(j,-n,k)$. Thus, we obtain
\begin{equation}  \label{selflong}
\Sigma
_{\alpha,\alpha^{\prime}}(\varepsilon)=\delta_{j,j^{\prime}}\delta_{k,k^{%
\prime}}
\left[\delta_{n,n^{\prime}}\Sigma^{d}(\varepsilon)+\delta_{n,-n^{\prime}}%
\Sigma^{o}(\varepsilon)\right].
\end{equation}
Splitting the self-energy into two parts
\beq
\Sigma_{\sf{tot}}^{\pm}\equiv\Sigma_{\sf{tot}%
}^{d}\pm\Sigma_{\sf{tot}}^{o}
\eeq
 we derive the positive contribution
\begin{equation}  \label{selfposit}
\Sigma_{\sf{tot}}^{+}(\varepsilon)=C|\om|\sum_{n=0}^{\infty}\sum_{k=0}^{\infty}\frac{%
(\varepsilon-\Sigma_{\sf{tot}}^{-})g(\varepsilon_{n,k})}{%
(\varepsilon-\Sigma_{\sf{tot}}^{+}) (\varepsilon-\Sigma_{\sf{tot}%
}^{-})-(\varepsilon_{n,k})^{2}}
\end{equation}
as well as the negative one
\begin{equation}  \label{selfnega}
\Sigma_{\sf{tot}}^{-}(\varepsilon)=C|\om|\sum_{n=1}^{\infty}\sum_{k=0}^{\infty}\frac{%
(\varepsilon-\Sigma_{\sf{tot}}^{+})g(\varepsilon_{n,k})}{%
(\varepsilon-\Sigma_{\sf{tot}}^{+}) (\varepsilon-\Sigma_{\sf{tot}%
}^{-})-(\varepsilon_{n,k})^{2}}
\end{equation}
where $C$ has the same form as for the short-range case, i.e.
(\ref{paramc}). These parts can be used to
derive the density of states for the long-range case. More precisely,
we obtain
\begin{equation}  \label{densitysta3}
\rho(\varepsilon)=-\frac{1}{\pi }\frac{1}{C|\om|} {\sf{Im}}%
[\Sigma_{\sf{tot}}^{+}(\varepsilon+i0)+\Sigma_{\sf{tot}%
}^{-}(\varepsilon+i0)]
\end{equation}
which reduces to that obtained in~\cite{Ando1} by switching off the
confining parameter $\ka$. With this, we finish the derivations of the
 tools needed to tackle different issues. In fact, we will see how
 the above results can be applied  to deal with the diamagnetism
of the present system and emphasize what makes the difference with respect to the case
without confinement.

\section{ Thermodynamic properties} 

Now we show the relevance of the above tools.
We focus on
the study of  the diamagnetism and proceed
in the standard way
evaluating different physical
quantities. More precisely,
we determine 
the magnetization,
number of fermions, and susceptibility
to describe the physical properties of the system.

\subsection{Thermodynamic quantities} 

We recall  useful definitions
of different thermodynamic quantities. The magnetization is given by
\begin{equation}  \label{magnet1}
{\cal M}=-\left(\frac{\partial \Omega}{\partial B}\right)_{\mu}
\end{equation}
where $\Omega(T,\mu,B)$ is the thermodynamic potential and $\mu$ is the
chemical potential. 

To determine the number of fermions we can use
one of two methods. The first one is
based on the definition
\begin{equation}  \label{concent}
{\cal N}=-\left(\frac{\partial \Omega}{\partial \mu}\right)_{B}
\end{equation}
to obtain the Maxwell relation
\begin{equation}  \label{relationMax}
\left(\frac{\partial {\cal M}}{\partial \mu}\right)_{B}=\left(\frac{\partial {\cal N}}{%
\partial B}\right)_{\mu}.
\end{equation}
On the other hand, 
in terms of the density of states $\rho$, we have
\begin{equation}  \label{densityN}
{\cal N}=\int_{-\infty}^{\infty}\rho(\varepsilon,B)f(\varepsilon)d\varepsilon,
\end{equation}
where the fermionic distribution is $f(\varepsilon)=1/\left(1+e^{(\varepsilon-\mu)/k_{B}T}\right)$.

From the above formulas, one can establish an interesting relation.
After a straightforward calculation, we get
\begin{equation}  \label{magnet2}
{\cal M}=\int_{-\infty}^{\infty}d\varepsilon
f(\varepsilon)\int_{-\infty}^{\varepsilon}d\varepsilon^{\prime} \frac{%
\partial\rho(\varepsilon^{\prime},B)}{\partial B}
\end{equation}
in terms of the density of states. Furthermore,
one can also obtain the magnetic susceptibility
\begin{equation}  \label{magnesuscep}
\chi=\frac{\partial {\cal M}}{\partial B}{\Bigg |}_{B=0}.
\end{equation}
The above quantities will be simplified much more
by considering the self-consistent Born approximation and
fixing the type of  disorder.

\subsection{Short-ranged disorder}

To calculate different quantities we specify the nature of disorder.
According to (\ref{densitysta2}) and (\ref{magnet2}),
the susceptibility in the self-consistent Born approximation for the
short-ranged disorder can be written as
\begin{equation}  \label{suscp1}
\chi=-\frac{1}{\pi}\frac{2}{\kappa C}\int_{-\infty}^{\infty}d\varepsilon
f(\varepsilon)\int_{-\infty}^{\varepsilon}d\varepsilon^{\prime}\  {\sf{Im}}
\frac{\partial ^{2}\Sigma_{\sf{tot}}(\varepsilon^{\prime},B)}{\partial
B^{2}}{\Bigg|}_{B=0}.
\end{equation}
It is convenient to introduce
\beq\label{shvar}
X=\varepsilon-\Sigma_{\sf{tot}}
\eeq
which allows us to rewrite
\begin{equation}  \label{functionxb}
\Sigma_{\sf{tot}} (\varepsilon,B)\equiv
\widetilde{\Sigma}_{\sf{tot}}(X,B)=\frac{C|\om|}{2}\sum_{n=-\infty}^{\infty}\sum_{k=0}^{\infty} \frac{g(%
\varepsilon_{n,k})} {X-\varepsilon_{n,k}}.
\end{equation}
According to the above equations
second  derivatives are needed.
The first-order derivative of $\Sigma_{\sf{tot}}$ with respect to $B$ gives
\begin{equation}  \label{derivself1}
\frac{\partial \Sigma_{\sf{tot}}(\varepsilon,B)}{\partial B}=\left[1-
\frac{\partial \widetilde{\Sigma}_{\sf{tot}}(X,B)}{\partial X}\right]%
^{-1} \frac{\partial \widetilde{\Sigma}_{\sf{tot}}(X,B)}{\partial B}
\end{equation}
which leads to
the second as
\begin{equation}  \label{derivself2}
\frac{\partial^{2} \Sigma_{\sf{tot}}}{\partial B^{2}}=\left[1-\frac{%
\partial \widetilde{\Sigma}_{\sf{tot}}}{\partial X}\right]^{-1}\left[%
\frac{\partial^{2} \widetilde{\Sigma}_{\sf{tot}}}{\partial B^{2}}-2\frac{%
\partial \widetilde{\Sigma}_{\sf{tot}}}{\partial X\partial B}\left(\frac{%
\partial \Sigma_{\sf{tot}}}{\partial B}\right)+\frac{\partial^{2}
\widetilde{\Sigma}_{\sf{tot}}}{\partial X^{2}}\left(\frac{\partial
\Sigma_{\sf{tot}}}{\partial B}\right)^{2}\right].
\end{equation}

One can expand (\ref{functionxb}) into a series in terms of
a  function $h$:
\begin{equation}  \label{functionth}
\widetilde{\Sigma}_{\sf{tot}}(X,B)=\frac{C}{2}\Delta t\left[\frac{1}{2%
}h(0) +\sum_{n=-\infty, \neq 0}^{\infty} \sum_{k=1}^{\infty}
h\left(2\left[n+k+1\right]\Delta t\right)\right]
\end{equation}
where  $\Delta t=|\om|=|l_{B}^{-2}-\kappa|$ and the
function has the form
\begin{equation}
h(t)=\frac{2Xg(\sqrt{t})}{%
X^{2}-t}.
\end{equation}
Note that, taking $\ka=0$ we recover the results obtained in \cite{Ando1}.
This tells us that those results have been  generalized to the present case
and we will see how they can be interpreted.

To go further, we introduce some relevant assumptions. If
the condition ${\sf{Im}}(X) \gg \sqrt{|\om|}$ is valid, one can simplify (\ref{functionth}) to
\begin{eqnarray}  \label{approxi1}
&& C\Delta t\left[\frac{1}{2} h(0)+\sum_{n=-\infty, \neq 0}^{\infty} \sum_{k=1}^{\infty}
h\left(2\left[n+k+1\right]\Delta t\right)\right]%
=
\nonumber \\
&& C\int_{0}^{\infty} \  h(t)dt
-C\frac{(\Delta t)^{2}}{12} \left[\ h^{\prime}(0)+
\frac{1}{2} h^{\prime}(\infty)\right]
\end{eqnarray}
where $\theta(\Delta t)^{3}$ is neglected. This leads to
\begin{equation}  \label{approxi2}
\widetilde{\Sigma}_{\sf{tot}}(X,B)-\widetilde{\Sigma}_{\sf{tot}%
}(X,0)=-\frac{C}{24}h^{\prime}(0)(\Delta t)^{2}
\end{equation}
which can be used to calculate the above derivatives. Otherwise, after neglecting all terms containing
a power of $X$ larger than $3$,
we obtain
\begin{eqnarray}
\frac{\partial \Sigma_{\sf{tot}}}{\partial B}{\Big |}_{B=0} &=& \left[1-\frac{%
\partial \widetilde{\Sigma}_{\sf{tot}}}{\partial X}\right]^{-1}\
\left[\frac{C}{6}\kappa e\frac{1}{X^{3}}\right]  \label{selfdiriv1}\\
\frac{\partial^{2} \Sigma_{\sf{tot}}}{\partial B^{2}}{\Big |}_{B=0}&=& \left[1-%
\frac{\partial \widetilde{\Sigma}_{\sf{tot}}}{\partial X}\right]%
^{-1}\ \left[-\frac{C}{6} e^{2}\frac{1}{X^{3}}\right] \label{selfdiriv2}.
\end{eqnarray}

With the help of (\ref{derivself1}-\ref{derivself2}) and (\ref{selfdiriv1}-\ref{selfdiriv2}),
the susceptibility (\ref{suscp1})  becomes
\begin{equation}  \label{suscp2}
\chi=\frac{e^{2}}{3\pi\ka}  \int_{-\infty}^{\infty}d\varepsilon f(\varepsilon)\
{\sf{Im}}\int_{-\infty}^{\varepsilon}d\varepsilon^{\prime}\left(1-\frac{%
\partial \widetilde{\Sigma}_{\sf{tot}}}{\partial X^{\prime}}\right)^{-1}\frac{1}{%
X^{\prime 3}}{\Bigg |}_{B=0}.
\end{equation}
By introducing 
\begin{equation}  \label{integra}
d\varepsilon^{\prime}=\left[1-\frac{\partial \widetilde{\Sigma}_{\sf{tot}%
}}{\partial X^{\prime}}\right]dX^{\prime}
\end{equation}
it is not hard to find
\begin{equation}  \label{suscp3}
\chi=-\frac{e^{2}}{6\pi\ka} \int_{-\infty}^{\infty}d\varepsilon f(\varepsilon)\
{\sf{Im}}\frac{1}{\left[\varepsilon-\Sigma_{\sf{tot}}(\varepsilon)\right]^{2}}%
{\Bigg|}_{B=0}.
\end{equation}

As far as the magnetization is concerned for the short-ranged disorder, one can use
(\ref{densitysta2}), (\ref{selfdiriv1}-\ref{selfdiriv2}) and (\ref{integra}). These
give
\begin{equation}\label{mangneti}
    {\cal M}=-\frac{e}{6\pi} \int_{-\infty}^{\infty}d\varepsilon f(\varepsilon)\
{\sf{Im}}\frac{1}{\left[\varepsilon-\Sigma_{\sf{tot}}(\varepsilon)\right]^{2}}.
\end{equation}
On the other hand, the number of fermions can be also formulated
as
\begin{equation}\label{numberparti}
     {\cal N}=-\frac{1}{\pi}\frac{2}{C|\om|}\int_{-\infty}^{\infty}d\varepsilon f(\varepsilon)\
{\sf{Im}}\Sigma_{\sf{tot}}(\varepsilon,B).
\end{equation}

Now we will see how the above results will be simplified by
making use of different considerations. In fact,
to underline what makes a difference with respect to other studies,
we consider the zero-field limit and determine the corresponding susceptibility. 
More precisely, in such limit (\ref{toexa2}) can be written as
\begin{equation}\label{zerofield}
    \Sigma_{\sf{tot}}(\varepsilon)= C\int_{0}^{\infty}tdt\ \frac{(\varepsilon-
    \Sigma_{\sf{tot}})g(t)}{(\varepsilon-\Sigma_{\sf{tot}})^{2}-t^{2}}-\frac{\kappa^{2}}{12}
    C\frac{1}{(\varepsilon-\Sigma_{\sf{tot}})^{3}}+\cdots
\end{equation}
which can be approximated by   assuming that
$\varepsilon\ll\varepsilon_{c}$. In this situation, we
 end up with
\begin{equation}\label{integration}
   \Sigma_{\sf{tot}}(\varepsilon)=-C(\varepsilon-\Sigma_{\sf{tot}})\log\left
    [-\frac{\varepsilon_{c}^{2}}{(\varepsilon-\Sigma_{\sf{tot}})^{2}}\right]-\frac{\kappa^{2}}{12}
    C\frac{1}{(\varepsilon-\Sigma_{\sf{tot}})^{3}}+\cdots.
\end{equation}
To further simplify the above form, one may consider the condition
 $\kappa\ll 1$ and choose an appropriate branch of the logarithm. Thus,
one gets
\begin{equation}\label{applog}
   \Sigma_{\sf{tot}}(\varepsilon)=\varepsilon-\varepsilon\left[2C f_{L}
    \left(-\frac{i\varepsilon}{2\Gamma C}\right)\right]^{-1}
\end{equation}
where $f_{L}(z)$ is the Lambert $C-$function (called also Omega function
which is the inverse function of $f(z)=ze^z$~\cite{corless}) and $ \Gamma$
is given by
\begin{equation}\label{gamma}
    \Gamma=\varepsilon_{c}\exp\left(-\frac{1}{2C}\right).
\end{equation}
In the region where the energy fulfills the constraint $|\varepsilon|\gg\Gamma$,
$\Sigma_{\sf{tot}}$ becomes
\begin{equation}\label{self1}
    \Sigma_{\sf{tot}}(\varepsilon+i0)\approx-2\varepsilon C\log\left|
    \frac{\varepsilon_{c}}{\varepsilon}\right|-i\pi \left|\varepsilon\right|C.
\end{equation}
Following from this expression, we can derive specific results and
offer different discussions. Let us split
(\ref{self1}) as
\begin{eqnarray}\label{spself1}
    \Sigma_{\sf{tot}}(\varepsilon+i0) \approx  -n_iu_i^2\left[2\varepsilon\log\left|
    \frac{\varepsilon_{c}}{\varepsilon}\right|+i\pi \left|\varepsilon\right|\right]
     - \frac{\ka^2}{4}\left[2\varepsilon\log\left|
    \frac{\varepsilon_{c}}{\varepsilon}\right|+i\pi \left|\varepsilon\right|\right].
\end{eqnarray}
The first term is similar to that obtained in~\cite{Ando1} and becomes exactly
the same if the constraint $\ka=0$ is taken into account. Accordingly,
we can interpret the second as a quantum correction.

To treat the susceptibility for  short-ranged disorder, we distinguish two cases.
In the first case  the energy is constrained by the condition $\varepsilon\ll\varepsilon_{c}$.
At zero temperature, we evaluate the integral to end up with
\begin{equation}\label{selfzertemp2}
    \chi(\varepsilon_{F})=-\frac{e^{2}}{3\pi}\frac{2C}{\Gamma}
    F\left(\frac{\varepsilon_{F}}{2\Gamma C}\right)
\end{equation}
where the function  $F(x)$ is given by
\begin{equation}\label{functionF}
    F(x)=-\frac{1}{x}{\mbox{Im}}\left[f_{L}(-ix)+\frac{1}{2}f_{L}^{2}(-ix)\right].
\end{equation}
Without confinement, this result reduces to
\begin{equation}\label{selfzertemp2k}
    \chi(\varepsilon_{F}){\Big|}_{\ka=0}=-\frac{e^{2}}{3\pi}\frac{2n_iu_i^2}{\Gamma|_{\ka=0}}
    F\left(\frac{\varepsilon_{F}}{2n_iu_i^2\Gamma|_{\ka=0}}\right)
\end{equation}
which is similar to that obtained in~\cite{Ando1}. Therefore,
(\ref{selfzertemp2}) is general in the sense that one can
change two
parameters, i.e. disorder $u_i$ and confinement $\ka$, to offer
different interpretations. In particular, we mention that
even when the disorder becomes smaller, the peak of (\ref{selfzertemp2})
does not become narrower. However, it happens
 when  $C$ becomes smaller.
On the other hand, by noticing that
$F(x)$ has its maximum at $x=0$ with
$F(0)=1$, we obtain 
\begin{equation}\label{susciptzer}
    \chi(\varepsilon_{F})=-\frac{e^{2}}{3\pi}\frac{2C}{\Gamma}.
\end{equation}

It is worthwhile to  see what happens in the case where $|\varepsilon|\gg\Gamma$.
Using  (\ref{self1})  the susceptibility becomes
\begin{equation}\label{self2}
    \chi(\varepsilon_{F})\approx -\frac{e^{2}}{3}\frac{C}{|\varepsilon_{F}|}.
\end{equation}
At this stage, we have different comments. First,
it is easy to see that if $C$ is constant then
(\ref{self2}) monotonically decreases as $|\varepsilon_{F}|$
increases. Second, let us write (\ref{self2}) as
\begin{equation}\label{sself2}
    \chi(\varepsilon_{F})\approx -\frac{e^{2}}{3}\frac{n_iu_i^2}{|\varepsilon_{F}|}
    -\frac{e^{2}}{12}\frac{\ka^2}{|\varepsilon_{F}|}
\end{equation}
which tells us that the first term is due to the disorder
and the second is a manifestation of the confinement. Clearly,
this can be interpreted as a quantum correction to the first
term. According to (\ref{sself2}), this conclusion disappears
if $\ka=0$. Furthermore,
in the limit $C\rightarrow0$, the susceptibility becomes a $\delta$-function
 \begin{equation}\label{suscipdelta}
     \chi(\varepsilon_{F})\approx -\frac{e^{2}}{6}\delta(\varepsilon_{F}).
\end{equation}
Obviously, this conclusion can not be reached in the clean limit $u_i\rightarrow 0$ as
in~\cite{Ando1}.

\subsection{Long-ranged disorder}

It is worthwhile to ask about
the susceptibility for the long-ranged disorder.
In  a similar way to the short-ranged case, we use
(\ref{densitysta3}) and (\ref{magnet2}) to obtain
\begin{equation}  \label{suscp4}
\chi=-\frac{2}{\pi}\frac{1}{\kappa C}\
\int_{-\infty}^{\infty}d\varepsilon
f(\varepsilon)\int_{-\infty}^{\varepsilon}d\varepsilon^{\prime}\ {\sf{Im}}%
\frac{1}{2}\frac{\partial^{2}}{\partial B^{2}}\left[\Sigma_{\sf{tot}%
}^{+}(\varepsilon^{\prime},B)+\Sigma_{\sf{tot}}^{-}(\varepsilon^{%
\prime},B)\right]{\Bigg |}_{B=0}.
\end{equation}
By analogy to (\ref{shvar}), we change the variable to
\beq
X^{\pm}=\varepsilon\pm\Sigma_{\sf{tot}}^{\pm}
\eeq
and define $\Sigma_{\sf{tot}}^{\pm}\equiv\widetilde{\Sigma}_{\sf{tot}%
}^{\pm}(X^{+},X^{-},B)$ as
\begin{equation}  \label{selfenergposi1}
\widetilde{\Sigma}_{\sf{tot}}^{+}\equiv C |\om|\sum_{n=0}^{\infty}\sum_{k=0}^{\infty}\frac{%
X^{-} g(\varepsilon_{n,k})}{X^{+}X^{-}-(\varepsilon_{n,k})^{2}}
\end{equation}
\begin{equation}  \label{selfenergposi2}
\widetilde{\Sigma}_{\sf{tot}}^{-}\equiv C|\om|\sum_{n=1}^{\infty}\sum_{k=0}^{\infty} \frac{%
X^{-} g(\varepsilon_{n,k})}{X^{+}X^{-}-(\varepsilon_{n,k})^{2}}.
\end{equation}
The derivatives of $\Sigma_{\sf{tot}}$ can be written in terms of $%
\widetilde{\Sigma}_{\sf{tot}}$ as
\begin{eqnarray}  
\frac{\partial \Sigma_{\sf{tot}}}{\partial B} &=& A_{ij}\frac{\partial
\widetilde{\Sigma}_{\sf{tot}}^{j}}{\partial B}
\label{deriv1}\\
\frac{\partial^{2}\Sigma_{\sf{tot}}^{i}}{\partial B^{2}} &=& A_{ij}\left[%
\frac{\partial^{2}\widetilde{\Sigma}_{\sf{tot}}^{j}}{\partial B^{2}}- 2%
\frac{\partial^{2}\widetilde{\Sigma}_{\sf{tot}}^{j}}{\partial X^{k}
\partial B}\frac{\partial \Sigma_{\sf{tot}}^{k}}{\partial B}+ \frac{%
\partial^{2}\widetilde{\Sigma}_{\sf{tot}}^{j}}{\partial X^{k}\partial
X^{l}}\frac{\partial \Sigma_{\sf{tot}}^{k}} {\partial B}\frac{\partial
\Sigma_{\sf{tot}}^{l}}{\partial B}\right]\label{deriv2}
\end{eqnarray}
where $i,j,l=\pm$ and repeated indices indicate summation. The involved matrix
elements are given by
\begin{equation}  \label{operateurA}
A_{ij}\equiv \left(\delta_{ij}+\frac{\partial \widetilde{\Sigma}_{\sf{tot}}^{i}}{\partial X^{j}}\right)^{-1}.
\end{equation}

One can calculate the derivatives of $\widetilde{\Sigma}^{\pm}$ at $B=0$, in a
similar way to the short-range case, and then obtain those for $\Sigma^{\pm}$
using  (\ref{deriv1}) and (\ref{deriv2}), to get
\begin{eqnarray}
  \frac{\partial}{\partial B}(\Sigma^{+}+\Sigma^{-}) &=& \left(1+\alpha+2\beta\right)^{-1}\
  \frac{\ka e}{X^{3}} \left(\frac{C}{6}\right)  \\
  \frac{\partial^{2}}{\partial B^{2}}(\Sigma^{+}+\Sigma^{-})&=&\left(1+\alpha+2\beta\right)^{-1} \
  \frac{ e^2}{X^{3}}\left(-\frac{C}{6}+\frac{1}{1-\alpha}\frac{C^{2}}{2}-
\frac{2\beta}{(1-\alpha)^{2}}\frac{C^{2}}{4}\right)
\end{eqnarray}
where $X\equiv\lim_{B\rightarrow0}X^{+}=\lim_{B\rightarrow0}X^{-}$%
 and the two parameters are
\begin{eqnarray}
\alpha &=& 2C\int_{0}^{\infty}tdt\ \frac{g(t)}{X^{2}-t^{2}}\label{alpha}\\
\beta &=& 2C\int_{0}^{\infty}tdt\ \frac{-X^{2}g(t)}{(X^{2}-t^{2})^{2}}\label{beta}.
\end{eqnarray}

We have now derived all ingredients to write the
expression for the susceptibility in this case:
\begin{equation}  \label{derivation}
\chi=-\frac{e^{2}}{6\pi\ka}  \int_{-\infty}^{\infty}d\varepsilon f(\varepsilon)%
\int_{X(-\infty)}^{X(\varepsilon)}dX^{\prime}\ {\sf{Im}}\frac{1}{X^{\prime3}}\left[%
-1+ 3C\left(\frac{1}{1-\alpha^{\prime}}-\frac{\beta^{\prime}}{%
(1-\alpha^{\prime})^{2}}\right)\right]{\Bigg |}_{B=0}
\end{equation}
where the integration over $\varepsilon^{\prime}$ has been replaced by
\begin{equation}
dX^{\prime}=\left(1+\alpha^{\prime}+2\beta^{\prime}\right)^{-1}d\varepsilon^{
\prime}.
\end{equation}

For the long-ranged disorder in the region $|\varepsilon|\ll\varepsilon_{c}$,
(\ref{alpha}) and (\ref{beta}) can be approximated  as
\begin{eqnarray}
    \alpha &\approx &- C\log\left(-\frac{\varepsilon_{c}^{2}}{X^{2}}\right)\label{alpha1}\\
    \beta &\approx & C \label{beta2}.
\end{eqnarray}
By substituting them in (\ref{derivation}) and performing the integration over
$X^{\prime}$, we see that the susceptibility is given by
\begin{equation}\label{suscipapprox}
    \chi=-\frac{e^{2}}{6\pi\ka} \int_{-\infty}^{\infty}d\varepsilon
    f(\varepsilon)\ {\sf
     Im} \frac{1}{X^{2}}\left[1-\frac{3C}{1+C\log\left(-\frac{\varepsilon_{c}^{2}}{X^{2}}\right)
     }\right]{\Bigg|}_{B=0}
\end{equation}
with $X=\varepsilon-\Sigma_{\sf{tot}}(\varepsilon)$.
This can be rewritten as
\begin{equation}\label{suscipapprox2}
    \chi=-\frac{e^{2}}{6\pi\ka} \int_{-\infty}^{\infty}d\varepsilon
    f(\varepsilon)\ {\sf
     Im} \frac{1}{X^{2}}{\Bigg|}_{B=0} + \frac{e^{2}}{2\pi\ka} C\int_{-\infty}^{\infty}d\varepsilon
    f(\varepsilon)\ {\sf Im} \frac{1}{X^2\left[1+ C\log\left(-\frac{\varepsilon_{c}^{2}}{X^{2}}\right)
\right]}{\Bigg|}_{B=0}.
\end{equation}
The first term is  the short-range contribution
(\ref{suscp3}). The second term can be regarded as
a contribution of
order of ${\cal O}(C)$, but this gives a minor effect since $C$ is assumed
to be small. When ${\cal O}(C)^{2}$ is neglected, the susceptibility becomes
just $1-3C$ times as large as in the case of short-ranged disorder. Accordingly
the integration of $\chi$ over $\varepsilon$ weakly depends on $C$, while
in $C\rightarrow0$ we again get (\ref{suscipdelta}).
Finally, note that taking  $\ka=0$ we end up with the
results obtained by Koshino and Ando in~\cite{Ando1}.

\section{Conclusion}

The present paper was devoted to give a complete solution to the confined Dirac
fermion system in the presence of a perpendicular magnetic field. Using a
similarity transformation we have formulated our problem in terms of the polar
coordinate representation that allows us to handle easily the exact relationship
between spinor components. One spinor component was obtained
by solving a second order differential equation while the other component was
obtained using the exact relationship (\ref{eq16}). It resulted in a complete solution
space made of 8 subspaces, which suggests that it is necessary to include all
components of this subspace in the computations of any physical quantity. A failure
to do so will result in erroneous conclusions.

Considering the disordered graphene, we formulated our solutions
to capture the impurity potential. After providing the necessary tools,
we used the Green function technique to determine
the self-energy. For this, two  cases are discussed,
which concern short and long-ranged disorder. These allowed us to
give a simplified form of the density of states
that was used to determine different
thermodynamical quantities.

For further studies, we distinguished between two cases. As far as the
short-ranged disorder is concerned, we further simplified the self-energy
and found a quantum correction to the  susceptibility
for Dirac fermions in disordered graphene. On the other hand, we noted that  $\ka=0$ allowed us to recover the
results of disordered graphene in the presence of a magnetic field~\cite{Ando1}.
Furthermore,
in the limit $C\rightarrow0$, the susceptibility becomes a
$\delta$-function (\ref{suscipdelta}).

Subsequently, we treated the long-ranged case
and stressed what makes a difference with respect to the former
one. By
comparing the results obtained in (\ref{suscp3}) and
(\ref{suscipapprox}), we noticed there is  an extra term of the
order of ${\cal O}(C)$, but this gives a minor effect since $C$ is assumed
to be small. When ${\cal O}(C)^{2}$ is neglected, the susceptibility becomes
just $1-3C$ times as large as in the short-ranged disorder. Accordingly
the integration of $\chi$ over $\varepsilon$ weakly depends on $C$, while
in the limit $C\rightarrow0$ we again get (\ref{suscipdelta}).

The present work can be extended to other cases. One may numerically check
the above results and compare them with those obtained before. On the other hand, an interesting
question arises about what happens for a variable magnetic field, in particular an exponential
variation of the field. This  question is under investigation.

\section*{Acknowledgment}

%
We thank E.B. Choubabi for discussions.
A part of this work was done during  AJ's visit to
the Max Planck Institute for the Physics of Complex Systems, Dresden. He would
like to thank the Institute for 
the warm hospitality.


\begin{thebibliography}{1}
\bibitem{explain} Any other choice for the pair of Pauli matrices can be obtained
from the present one by a unitary transformation, hence leaving
the physics of the problem unaltered.
\end{thebibliography}
\end{document}